# The prospects of quantum computing in computational molecular biology

Carlos Outeiral[1,2]  |  Martin Strahm[3]  |  Jiye Shi[4]  |  Garrett M. Morris[1]  |
Simon C. Benjamin[2]  |  Charlotte M. Deane[1]

[1]Department of Statistics, University of Oxford, Oxford, UK

[2]Department of Materials, University of Oxford, Oxford, UK

[3]Pharma Research and Early Development, F. Hoffmann-La Roche, Basel, Switzerland

[4]Computer-Aided Drug Design, UCB Pharma, Slough, UK

**Correspondence**
Charlotte M. Deane, Department of Statistics, University of Oxford, 24–29 St Giles', Oxford OX1 3LB, UK.
Email: deane@stats.ox.ac.uk

**Funding information**
Engineering and Physical Sciences Research Council, Grant/Award Numbers: EP/L016044/1, EP/M013243/1, EP/S024093/1, EP/T001062/1

**Abstract**

Quantum computers can in principle solve certain problems exponentially more quickly than their classical counterparts. We have not yet reached the advent of useful quantum computation, but when we do, it will affect nearly all scientific disciplines. In this review, we examine how current quantum algorithms could revolutionize computational biology and bioinformatics. There are potential benefits across the entire field, from the ability to process vast amounts of information and run machine learning algorithms far more efficiently, to algorithms for quantum simulation that are poised to improve computational calculations in drug discovery, to quantum algorithms for optimization that may advance fields from protein structure prediction to network analysis. However, these exciting prospects are susceptible to "hype," and it is also important to recognize the caveats and challenges in this new technology. Our aim is to introduce the promise and limitations of emerging quantum computing technologies in the areas of computational molecular biology and bioinformatics.

This article is categorized under:
  Structure and Mechanism > Computational Biochemistry and Biophysics
  Data Science > Computer Algorithms and Programming
  Electronic Structure Theory > Ab Initio Electronic Structure Methods

## 1 | INTRODUCTION

Since the advent of modern computing, algorithms and mathematical models have been used to help solve biological problems, from exploring the complexity of the human genome to modeling the behavior of biomolecules. Computational methods are now routinely used to inform and extract information from biological experiments, as well as predict the behavior of biological entities and systems. In fact, 10 of the 25 most highly cited scientific papers relate to algorithms used in biology [1], including quantum simulation [2–5], sequence alignment, [6–8], computational genetics [9] and X-ray diffraction data processing [10,11]. Despite such progress, many challenges in biology remain computationally infeasible. The best algorithms for problems like predicting the folding of a protein, calculating the binding affinity of a ligand for a macromolecule, or finding optimal large-scale genomic alignments require computational resources that are beyond even the most powerful supercomputers of our era.

The solution to these challenges may lie in a paradigm shift in computing. In the 1980s, Richard Feynman [12] and, independently, Yuri Manin [13] proposed using quantum mechanical effects to build a new, more powerful generation of computers. Quantum theory has proved to be a highly successful description of physical reality, and has led, since its introduction in the early 20th century, to advances such as lasers, transistors and semiconductor microprocessors. A *quantum* computer would enable more effective algorithms by introducing operations that are not possible in classical machines. Quantum processors do not work faster than classical computers, but operate in a fundamentally different way, achieving unprecedented speedups by avoiding unnecessary computation. For example, computing the full electronic wavefunction of an average drug molecule numerically is expected to take longer than the age of the universe on any current supercomputer using conventional algorithms [14], while even a modest-sized quantum computer may be able to solve this in a timescale of days. Motivated by this promise of *quantum advantage*, the quest to build a quantum processor is ongoing. Unfortunately, the technical difficulties in manufacturing, controlling and protecting quantum systems from noise are staggering, and the first prototypes have only appeared in the last decade.

The technical challenges of building quantum computers have not stopped the development of quantum algorithms. Even in the absence of hardware, algorithms can be analyzed mathematically, and the emergence of high-performance simulators of quantum computers, as well as early prototypes in the last few years, have enabled further investigation. Some of these algorithms have already shown promise for applications in biology. For example, the quantum phase estimation algorithm allows exponentially faster eigenvalue calculation [15], which can be used to understand large scale correlation between portions of a protein or determine centrality in a biological network. The Harrow–Hassidim–Lloyd (HHL) quantum algorithm [16] can solve certain linear systems exponentially faster than any known classical algorithm, and so could power statistical learning techniques that can be trained more quickly and manage more data. Quantum optimization algorithms have potential in the field of protein folding and conformer sampling, and across problems that involve finding minima or maxima [17]. Finally, technology to simulate quantum systems has been developed which promises accurate predictions of drug–receptor interaction [18], or insight into chemical mechanisms such as photosynthesis [19]. Quantum computing has the potential to change the way we do biology much as classical computing did.

The recent claims of quantum supremacy by Google [20], although challenged by IBM [21] indicate that the era of quantum computing will soon be with us. Early processors leveraging quantum effects to perform classically impossible calculations are expected within the next decade [22]. In this context, the objective of this article is to set the scene as to the current areas where quantum computing shows promise for computational biology. Several recent reviews have analyzed the potential impact of quantum computing in a variety of fields, including machine learning [23–25], quantum chemistry [26–28] and drug discovery [18]. There has also recently been a report from a NIMH workshop on quantum computing in the biological sciences [29]. In this review, we will first provide a brief outline of what is meant by quantum computing, and a short introduction to the principles of quantum information processing. Then we will discuss three main areas of computational biology where quantum computing has already shown promising algorithmic developments: statistical methods, electronic structure calculations, and optimization. Due to the limitations of space, we are forced to leave aside some topics, for example, string algorithms that could impact sequence analysis [30], medical imaging algorithms [31], numerical algorithms for differential equations [32,33] and other mathematical problems [34], or methods to analyze biological networks [35–38]. Finally, we will discuss the potential impact of quantum computing to computational biology in the medium- and long-term.

## 2 | QUANTUM INFORMATION PROCESSING

Quantum computers promise to solve biologically relevant problems, like predicting protein–ligand interactions or understanding coevolution of amino acids in proteins, exponentially faster than any current computer. This paradigm shift, however, requires a fundamental change in our thinking: quantum computers are very different from their classical analogues. The physical phenomena that underpin quantum advantage are often counter-intuitive and defy common sense, and utilizing a quantum processor requires fundamental changes in the way we understand programming. In this section, we introduce the principles of quantum information and how it is processed to perform computation.

We explain how information works differently in a quantum system, where it is stored in qubits, and how this information may be manipulated using quantum gates. Like the variables and functions of a programming language, qubits and quantum gates establish the basic components of any algorithm. We also discuss why building a quantum computer is technically challenging, and what may be achieved with the early prototypes that are expected in the coming years. This introduction will cover only the basic requirements; for a comprehensive introduction see Nielsen and Chuang [39].

## 2.1 | The elements of quantum algorithms

### 2.1.1 | Quantum information: Introducing the qubit

The first challenge in presenting quantum computing is introducing how it handles information. In a quantum processor, information is generally stored in *qubits*, which are the quantum analogue of classical bits. A qubit is a physical system, like an ion confined to a magnetic field [40] or a polarized photon [41], but it is often spoken of in an abstract fashion. Much like Schrödinger's cat, a qubit may not only adopt the states 0 or 1, but also any possible combination of both states. When the qubit is observed directly, it will no longer be in a superposition, but will *collapse* to one of the possible states, in the same way that Schrödinger's cat is either dead or alive after the box is opened [42]. More importantly, when several qubits are combined, they may become correlated, and interactions with any single one of them have subtle implications for the entire collective state. As we will see, the phenomenon of correlation between multiple qubits, known as quantum entanglement, is a fundamental resource of quantum computation.

In classical information, the fundamental unit of information is a *bit*, a system with two identifiable states often named 0 and 1. The quantum analogue, the *qubit*, is a two-state system whose states are labeled $|0\rangle$ and $|1\rangle$. We employ Dirac notation, where $|\cdot\rangle$ identifies a quantum state. The primary difference between classical and quantum information is that a qubit can be in any *superposition* of the states $|0\rangle$ and $|1\rangle$:

$$|\psi\rangle = \alpha|0\rangle + \beta|1\rangle \qquad \alpha,\beta \in \mathbb{C} \qquad |\alpha|^2 + |\beta|^2 = 1. \tag{1}$$

The complex coefficients $\alpha$ and $\beta$ are known as the *amplitudes* of the respective states, and they are related to another key aspect of quantum mechanics: the effect of physical measurement. Since qubits are physical systems, it is always possible to conceive a protocol to measure their state. If, for example, the states $|0\rangle$ and $|1\rangle$ correspond to the spin-down and spin-up states of an electron in a magnetic field, then measurement of the state of the qubit in this basis is just a measurement of the energy of the system. The postulates of quantum mechanics dictate that if the system is in a superposition of the possible measurement outcomes, then the act of measurement must alter the state itself. The system will *collapse* to the measured state; measurement thus destroys the information carried by the amplitudes in a qubit [39].

The important consequences for computing emerge when we consider systems of multiple qubits, which can experience *quantum entanglement*. Entanglement is a phenomenon where a group of qubits are correlated, and any operation on one of these qubits will affect the collective state of all of them. The canonical example of quantum entanglement is the Einstein–Podolsky–Rosen paradox presented in 1935 [43]. Consider a system of two qubits where, since each of the individual qubits may adopt any superposition of the states $\{|0\rangle, |1\rangle\}$, the composite system can adopt any superposition of the states $\{|00\rangle, |01\rangle, |10\rangle, |11\rangle\}$ (and by extension, an $N$-qubit system can adopt any of the $2^N$ binary strings, from $|0...0\rangle$ to $|1...1\rangle$). One of the possible superpositions of the system are the so-called *Bell states*, one of which has the following form:

$$|\psi\rangle = \frac{1}{\sqrt{2}}(|10\rangle + |01\rangle) \tag{2}$$

If we perform a measurement on the first qubit, we will only be able to observe $|0\rangle$ or $|1\rangle$, each of them with probability 1/2. This introduces no changes with respect to the single-qubit case. If the outcome for the first qubit was $|0\rangle$, the system will have collapsed to the $|01\rangle$ system, and therefore any measurement on the second qubit will yield $|1\rangle$ with a probability of 1; likewise, if the result of the first measurement was $|1\rangle$, the measurement on the second qubit will yield $|0\rangle$. An operation (in this case, measurement with outcome "0") applied to the first qubit affects the outcomes that can be seen by a subsequent measurement of the second qubit.

The existence of entanglement is fundamental for useful quantum computation. It has been proved that any quantum algorithm that does not employ entanglement can be enacted in a classical computer with no significant difference in speed [44,45]. The intuitive reason is the magnitude of information that a quantum computer can manage. If an $N$-qubit system is not entangled, the $2^N$ amplitudes of its state may be described by the amplitudes of each single-qubit state, that is, $2N$ amplitudes. However, if the system is entangled, all the amplitudes will be independent and the qubit register will form a $2^N$-dimensional vector. The ability of quantum computers to apparently manipulate vast amounts of information with few operations is one of the cornerstones of quantum algorithms, and underpins their capacity to solve problems exponentially faster.

### 2.1.2 | Quantum gates

The information stored in qubits is manipulated using special operations known as *quantum gates*. Quantum gates are physical operations, like a laser pulse targeted to an ion qubit, or a set of mirrors and beamsplitters through which a photonic qubit must move [39], but they are often considered as abstract operations. The postulates of quantum mechanics impose some strict conditions on the nature of quantum gates within closed systems, which allows them to be represented as *unitary* matrices, which are linear operations that preserve the normalization of a quantum system. In particular, a quantum gate applied to an entangled register of $N$ qubits is equivalent to multiplying a $2^N \times 2^N$ matrix times a $2^N$-entry vector. The ability of a quantum computer to store and perform calculations on vast amounts of information—of the order of $2^N$—by manipulating a few elements—of order $N$—forms the basis of its ability to provide a potentially exponential advantage over classical computers.

In essence, a quantum gate is any allowed operation on a system of qubits. The postulates of quantum mechanics impose two strict constraints on the form of quantum gates. In the first place, quantum operators are *linear*. Linearity is a mathematical condition that nonetheless has profound implications for the physics of quantum systems—and therefore how they can be used for computing. If a linear operator $\hat{O}$ is applied to a superposition of states, the result is a superposition of the individual states affected by the operator. In a qubit, this means:

$$\hat{O}(\alpha|0\rangle + \beta|1\rangle) = \alpha(\hat{O}|0\rangle) + \beta(\hat{O}|1\rangle) = \alpha'|0\rangle + \beta'|1\rangle. \tag{3}$$

Linear operators can be represented as *matrices*, which are simply tables representing the effect of the linear operator on every basis state. In Figure 1c,d we show the matrix representation of one two-qubit and two one-qubit gates. However, not every matrix represents a valid quantum gate. We expect that a quantum gate applied to a collection of qubits will yield another valid collection of qubits, in particular one that is normalized (e.g., in Equation (3), we expect $|\alpha'|^2 + |\beta'|^2 = 1$). This condition holds only if the matrix representing a quantum gate is unitary, that is, $U^\dagger U = UU^\dagger = I$, where $U^\dagger$ is the matrix $U$ where rows and columns have been interchanged, and every complex number has been conjugated (meaning that every imaginary element acquires a negative factor). An arbitrary $2^N \times 2^N$ unitary matrix represents a completely valid $N$-qubit quantum gate.

In classical computing there is only one nontrivial gate for a single bit: the NOT gate, that transforms 0 into 1 and vice versa. In quantum computing, there is an infinite number of $2 \times 2$ unitary matrices, and any of them is a possible one-qubit quantum gate. One of the early successes of quantum computing was the discovery that this huge number of possibilities can be implemented with a set of *universal* gates affecting one and two qubits [47]. In other words, given an arbitrary quantum gate, there is a circuit built from one- and two-qubit gates that can enact it to arbitrary precision. Unfortunately, this does not mean that the approximation is efficient. Most quantum gates can only be approximated with an exponential number of gates from our universal set [39]. Even if these gates can be used to solve useful problems, implementing them will take exponentially large amounts of time, and may obliterate any quantum advantage.

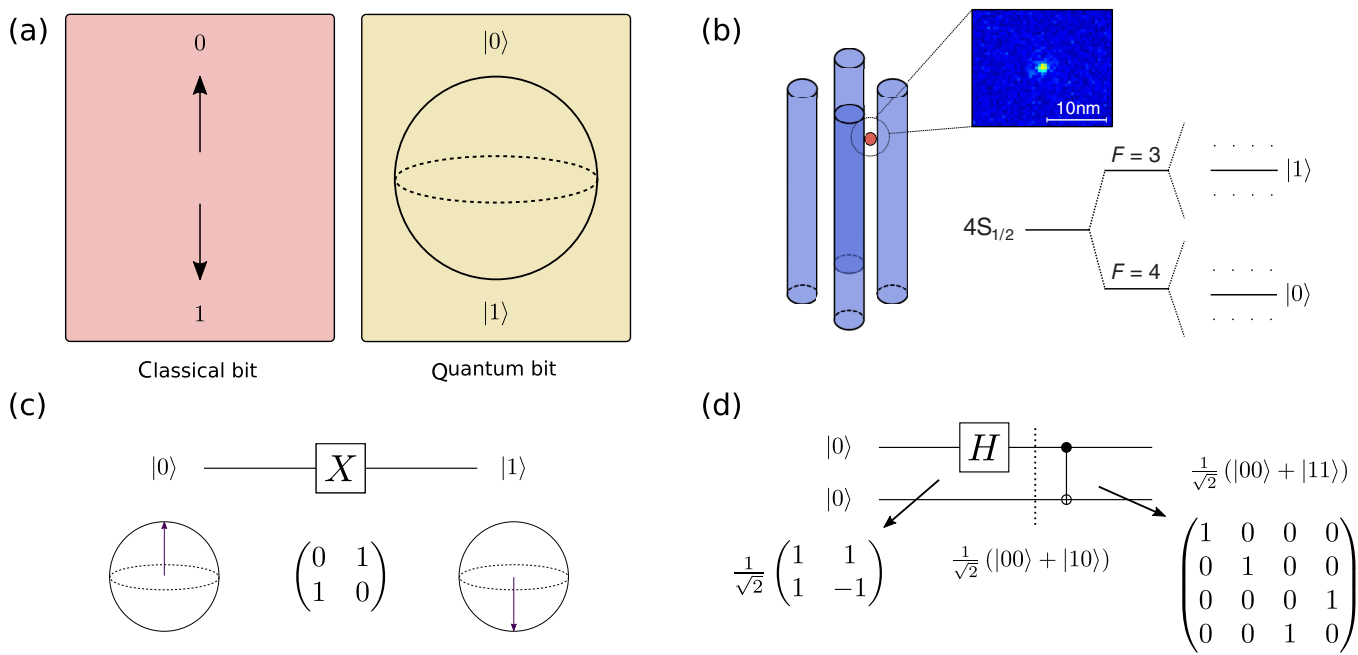

**FIGURE 1** (a) Comparison between a classical bit and a quantum bit or "qubit." While the classical bit can only take one of two states, 0 or 1, the quantum bit can take any state of the form $|\psi\rangle = \cos\frac{\theta}{2}|0\rangle + e^{i\phi}\sin\frac{\theta}{2}|1\rangle$. Single qubits are often depicted using the Bloch sphere representation, where $\theta$ and $\phi$ are understood as the azimuthal and polar angles in a sphere of unit radius. (b) Scheme of an ion trap qubit, one of the most common approaches to experimental quantum computing. An ion (often $^{43}Ca^+$) is confined in high vacuum using electromagnetic fields, and is subjected to a strong magnetic field. The hyperfine structure levels are split according to the Zeeman effect, and two selected levels are chosen as the states $|0\rangle$ and $|1\rangle$. Quantum gates are implemented by appropriate laser pulses, often involving other electronic levels. This diagram has been adapted from [46]. (c) Diagram of a quantum circuit implementing the $X$ or quantum-NOT gate. We show the matrix representation, and the change in the Bloch sphere. (d) Quantum circuit to generate a Bell state $\frac{1}{\sqrt{2}}(|00\rangle + |11\rangle)$ using the Hadamard $H$ gate and the controlled-NOT gate. The dotted line in the middle of the circuit indicates the state after applying the Hadamard gate

## 2.2 | Quantum hardware

Quantum algorithms can only solve interesting problems if they are run in appropriate quantum hardware. There are many competing proposals to build a quantum processor, including trapped ions [40], superconducting circuits [48] and photonic devices [41]. However, all of them face a common challenge: errors during computation, which can ruin the computational process. One of the cornerstones of quantum computing is the finding that these errors can be suppressed by using quantum error correcting codes. Unfortunately, these codes also require a very large increase in the number of qubits, so significant engineering advances are necessary before fault-tolerance can be achieved.

There are many sources of error that can affect the quantum processor. For example, coupling of a qubit to its environment can cause the system to collapse to one of its classical states, a process known as *decoherence*. Small fluctuations can transform the quantum gate we hope to apply into a similar one, producing a different output than expected, and the imperfect control mechanisms will always introduce a certain magnitude of error. The least error-prone gates to date have been reported in a trapped ion processor, with error rates of one part in $10^6$ for single-qubit gates and errors of 0.1% for two-qubit gates [46,49]. For comparison, the recent work on quantum supremacy by Google reported fidelities of 0.1% for single-qubit gates and 0.3% for two-qubit gates in a superconducting processor [20]. Given that a single gate failure could corrupt a sensitive calculation, it is not difficult to see that error propagation can render the computation meaningless after only a small sequence of gates.

One of the cornerstones of quantum computing is the development of quantum error correcting codes. In the 1990s, several groups proved that these codes can achieve *fault-tolerant computation*, as long as the error rates of the gates are under a particular threshold that depends on the code [50–53]. One of the most popular approaches, the *surface code*, can work with error rates approaching 1% [54]. Unfortunately, quantum error correcting codes require using a large number of real, *physical qubits* to encode an abstract *logical qubit* that is used for computation, and this overhead becomes worse as the error rate increases. For example, the quantum algorithm for prime number factorization [55] could, in noiseless conditions, factorize a 2,000-bit number employing approximately 4,000 qubits and, assuming a 16 GHz gate rate, about a day to run. Assuming a 0.1% error rate, this same algorithm, using the surface code to correct environmental errors, would require several million qubits and a similar amount of time [54]. Given that the current record for a controllable, programmable quantum processor is 53 qubits [20], there is still a long way to go.

Many groups have undertaken efforts to develop algorithms that can be run in the so-called noisy intermediate-scale quantum processors [22]. Variational algorithms, for example, combine a classical computer with a small quantum processor, running a large number of short quantum computations that can be implemented before the noise takes over. These algorithms often use a parametrized quantum circuit that performs a particularly difficult task, and employ the classical computer to optimize the parameters. An enabling technique is the area of error mitigation which has attempted to, instead of achieving fault-tolerance, reduce errors as much as possible with minimum effort to run larger circuits. There are a number of approaches that include applying extra operations to discard computational runs with errors [56] or manipulating the error rate to extrapolate to the correct result [57,58]. Although major applications will require very large, fault-tolerant quantum computers, it is expected that near-term devices available within the next decade will be able to solve useful problems [22].

# 3 | STATISTICAL METHODS AND MACHINE LEARNING

In computational biology, where the objective is often to assimilate vast amounts of data, statistical methods and machine learning are key techniques. For example, in genomics, annotation of gene information has made extensive use of hidden Markov models (HMMs) [59]; in drug discovery, a vast array of statistical models have been developed to estimate molecular properties, or to predict if a ligand will bind to a protein [60]; and, in structural biology, deep neural networks have been used to both predict contacts [61], secondary structure [62] and most recently 3D protein structures [63]. Training and developing such models is often computationally intensive. A major catalyst of the recent advancements in machine learning was the realization that general purpose graphical processing units (GPUs) could significantly accelerate the training procedure. By providing exponentially faster algorithms to train machine learning models, quantum computing may provide a similar stimulus to scientific applications.

In this section, we explain how quantum computing can accelerate numerous statistical learning methods. Our introduction is aimed at developments that are likely to be directly applicable in biology; for a wider treatment of quantum machine learning, see [23–25].

## 3.1 | Advantages and shortcomings of quantum machine learning

We first consider the kind of advantages that a quantum computer presents for machine learning. Save for idealized examples, quantum computers can learn no more information than a classical computer from the same training set [64]. However, they can in principle be far faster and able to handle far more data than their classical counterparts. As an example, the human genome contains 3 billion base pairs, which can be stored in $1.2 \times 10^{10}$ classical bits—approximately 1.5 gigabytes. A register of $N$ qubits involves $2^N$ amplitudes, which can each represent a classical bit, by setting the $k$th amplitude to 0 or 1 with an appropriate normalization factor. Therefore, a human genome could be stored in ∼34 qubits. Although this information could not be extracted from the quantum computer, as long as the state can be efficiently prepared (a considerable assumption, as later discussed) it would be possible to execute certain machine learning algorithms on it. More importantly, doubling the size of the register to 68 qubits leaves just about enough space to store the complete genome of *every* alive human in the world. The representation and analysis of such massive amounts of data would be well within the capabilities of even a small fault-tolerant quantum computer.

The operations to process this information could also run exponentially faster. For example, multiple machine learning algorithms are limited by the lengthy inversion of the covariance matrix, with a cost $\mathcal{O}(N^3)$ on the dimension of the matrix. However, the algorithm proposed by Harrow, Hassidim and Lloyd [16] allows the inversion of a matrix in $\mathcal{O}(\log N)$ within some conditions. In Table 1 we display some of the most common quantum machine learning algorithms and their theoretical quantum speedups. The key insight is that, unlike the acceleration provided by GPUs, which speed up computation through massive parallelism, quantum algorithms have a complexity advantage:

**TABLE 1** Overview of the main quantum machine learning algorithms that have been reported in the literature, and their time complexities

| Algorithm | Classical | Quantum | QRAM | References |
| --- | --- | --- | --- | --- |
| Linear regression | $\mathcal{O}(N)$ | $\mathcal{O}(\log N)^*$ | Yes | [65–68] |
| Gaussian process regression | $\mathcal{O}(N^3)$ | $\mathcal{O}(\log N)^\dagger$ | Yes | [69,70] |
| Decision trees | $\mathcal{O}(N \log N)$ | Unclear | No | [71] |
| Ensemble methods | $\mathcal{O}(N)$ | $\mathcal{O}(\sqrt{N})$ | No | [72–74] |
| Support vector machines | $\approx \mathcal{O}(N^2)\text{-}\mathcal{O}(N^3)$ | $\mathcal{O}(\log N)$ | Yes | [75–77] |
| Hidden Markov models | $\mathcal{O}(N)$ | Unclear | No | [78,79] |
| Bayesian networks | $\mathcal{O}(N)$ | $\mathcal{O}(\sqrt{N})$ | No | [80,81] |
| Graphical models | $\mathcal{O}(N)$ | Unclear | No | [82] |
| $k$-Means clustering | $\mathcal{O}(kN)$ | $\mathcal{O}(\log kN)$ | Yes | [83–85] |
| Principal component analysys | $\mathcal{O}(N)$ | $\mathcal{O}(\log N)$ | No | [86] |
| Persistent homology | $\mathcal{O}(\exp N)$ | $\mathcal{O}(N^5)$ | No | [87] |
| Gaussian mixture models | $\mathcal{O}(\log N)$ | $\mathcal{O}(\text{polylog } N)$ | Yes | [88,89] |
| Variational autoencoder | $\mathcal{O}(\exp N)$ | Unclear | No | [90] |
| Multilayer perceptrons | $\mathcal{O}(N)$ | Unclear | No | [91–95] |
| Convolutional neural networks | $\mathcal{O}(N)$ | $\mathcal{O}(\log N)$ | No | [96] |
| Bayesian deep learning | $\mathcal{O}(N)$ | $\mathcal{O}(\sqrt{N})$ | No | [97] |
| Generative adversarial networks | $\mathcal{O}(N)$ | $\mathcal{O}(\text{polylog } N)$ | No | [98–100] |
| Boltzmann machines | $\mathcal{O}(N)$ | $\mathcal{O}(\sqrt{N})$ | No | [101–105] |
| Reinforcement learning | $\mathcal{O}(N)$ | $\mathcal{O}(\sqrt{N})$ | No | [106,107] |

*Note*: We consider only the influence of the training set size N, however there are other factors (number of features, dimensionality, etc.) that can affect time complexity. When the exact solution of a problem is intractable, we report the most common heuristic (e. g. the Baum-Welch algorithm for HMMs). This table does not take into consideration other quantum advantages, like higher representational power or generalization ability. *: The cost is $\mathcal{O}(\text{polylog } N)$ if the coefficients are to be retrieved. †: This does not take into account the condition number of the covariance matrix, which can make the speedup only polynomial. A subset of the rows in this table were identified in the earlier work of Biamonte et al. [23].
Abbreviation: QRAM, quantum random access memory.

fewer operations are required to achieve the same task. In some cases, particularly when there exists an exponential speedup, moderately sized quantum computers could tackle training problems only available to the largest classical supercomputers available today.

Improved storage and processing of data has secondary benefits. One of the strengths of neural networks is their ability to find a concise representation of the data to which they are exposed [108]. Since quantum information is more general than classical information (after all, the states of a classical bit are subsumed by the eigenstates |0⟩ and |1⟩ or a qubit), it is possible that quantum machine learning models may be better at assimilating information than classical models. On the other hand, quantum algorithms with logarithmic-time complexity also allow for enhanced data privacy [83]. Since it takes $\mathcal{O}(\log N)$ to train a model, but $\mathcal{O}(N)$ calls to reconstruct the matrix, for a sufficiently large dataset it is impossible to recover a significant portion of the information, while it is still possible to train a model efficiently. In the context of biomedical research, this might encourage data sharing while ensuring confidentiality.

Unfortunately, although on paper quantum machine learning algorithms can vastly outperform classical analogues, practical difficulties still remain. Quantum algorithms are often posed as subroutines, which transform an input into an output. The problems appear precisely in these two steps: how to prepare an adequate input, and how to extract the information from the output [109]. Suppose for example that we employ the HHL algorithm [16] to solve a linear system of the form $A\vec{x} = \vec{b}$. At the end of the subroutine, the algorithm will yield a register of qubits in the following state:

$$|x\rangle = \frac{\sum_{j=1}^{N} \frac{\beta_j}{\lambda_j} |u_j\rangle}{\left\| \sum_{j=1}^{N} \frac{\beta_j}{\lambda_j} |u_j\rangle \right\|}. \quad (4)$$

Here $u_j$ and $\lambda_j$ are the eigenvectors and eigenvalues of $A$, and $\beta_j$ is the $j$th coefficient of $\vec{b}$ expressed in terms of the eigenvectors of A, and the denominator is just a normalisation constant. One can see this corresponds to the coefficients of $\vec{x}$, which are stored in the amplitudes of the different states as $x_j = \beta_j \lambda_j^{-1}$, and are not directly accessible. Measurement of the qubit register will cause it to collapse to one of the eigenvector states, and re-estimating the amplitudes of every $x_j$ requires $\mathcal{O}(2^N)$ measurements, which outweighs any advantage of the quantum algorithm in the first place. The HHL, and many other algorithms, are only useful to compute a *global* property of the solution, for example an expected value. In other words, the HHL cannot provide the solution of a system of equations or invert a matrix in logarithmic time, unless we are only interested in global properties of the solution that can be obtained with few measurements of physical observables. This limits the usage of some subroutines, but as we will show later, there have been many proposals to sidestep this problem.

The input of information to a quantum computer is a much more serious problem. Most quantum machine learning algorithms assume the quantum computer has access to the dataset as a superposition state; for example, there is a register of qubits that is, in a state of the form:

$$|\psi\rangle = \frac{\sum_j \alpha_j |\text{bin}(j)\rangle}{\left\| \sum_j \alpha_j |\text{bin}(j)\rangle \right\|}. \quad (5)$$

Here $|\text{bin}(j)\rangle$ is a state that acts as an index, and $\alpha_j$ is the corresponding amplitude. This can be used, for example, to store the elements of a vector or a matrix. In principle, there exists a quantum circuit that can prepare this state by acting on, say, the $|0...0\rangle$ state. However, implementing it can be very challenging, since we expect that approximating a random quantum state would be exponentially hard and wipe out any conceivable quantum advantage [39]. Most quantum algorithms have assumed access to a quantum random access memory (QRAM) [110], which is a black-box device able to construct this superposition. Although some blueprints have been proposed [111,112], there is not, to our knowledge, yet a working device. Moreover, even if such a device were available, there is no guarantee that it will not introduce bottlenecks that would outweigh the advantages of a quantum algorithm. For example, a recent circuit-based proposal for QRAM [113] shows an unavoidable linear cost on the number of states that would outweigh any logarithmic-time algorithm. Finally, even if the QRAM introduced no further overhead, it would still be necessary to pre-process data classically—for the genome example, access to 12 exabytes of classical storage would be required.

Finally, we must stress that quantum machine learning algorithms are not free of one of the most common problems in practical applications: scarcity of appropriate data. The availability of large amounts of data has been critical for the success of many practical applications of AI in molecular science, for example in de novo molecule design [114]. However, the power of quantum algorithms may prove useful as scientific and technological developments, such as the emergence of self-driving laboratories [115], provide more and more data.

Quantum machine learning could change how we process and analyze biological data. Unfortunately, the current practical challenges are sizeable.

## 3.2 | Quantum machine learning algorithms

### 3.2.1 | Unsupervised learning

Unsupervised learning encompasses several techniques to extract information from unlabelled datasets. In biology, where next-generation sequencing and large international collaborations have catalyzed data collection, these techniques have found widespread use, for example to identify relationships between families of biomolecules, [116] or annotate genomes [117].

One of the most popular unsupervised learning algorithms is principal component analysis (PCA), which attempts to reduce the dimensionality of the data by finding the linear combinations of features that maximize the variance [118]. This technique is widely used in all kinds of high-dimensionality datasets, including RNA microarray and mass spectrometry data [119]. A quantum algorithm for performing PCA was proposed by Lloyd et al. [86]. In essence, this algorithm builds the covariance matrix of the data in the quantum computer, and uses a subroutine known as *quantum phase estimation* [15] to compute the eigenvectors in logarithmic time. The output of the algorithm is a superposition state of the form:

$$\sum_{j=0}^{N} r_j |\eta_j\rangle. \tag{6}$$

Here $|\eta_j\rangle$ is the $j$th principal component, and $r_j$ is the corresponding eigenvalue. Since PCA is interested in large eigenvalues, which are the main components of the variance, measurement of the final state will yield a suitable principal component with high probability. Repetition of the algorithm several times will provide a set of principal components. This procedure is able to reduce the dimensionality of the vast amounts of information that can be stored in a quantum computer.

A quantum algorithm has also been proposed for a particular method in topological data analysis, namely *persistent homology* [87]. Topological data analysis tries to extract information by exploiting topological properties in the geometry of datasets; it has been used, for example, in the study of aggregation data [120] and network analysis [121]. Unfortunately, the best classical algorithms have a cost that is exponential in the dimensionality of the problem, which has limited its application. The algorithm by Lloyd et al. also makes use of the quantum phase estimation subroutine to diagonalize a matrix exponentially faster, reaching a $\mathcal{O}(N^5)$ complexity. The availability of an efficient algorithm to perform topological analysis might catalyze its application to analyze biological problems.

### 3.2.2 | Supervised learning

Supervised learning refers to a set of techniques that can make predictions from labeled data. The objective is to construct a model that can classify or predict the properties of unseen examples. Supervised learning has been used widely in biology, for problems as diverse as predicting the binding affinity of a ligand to a protein [122] and computer-aided disease diagnosis [123]. We discuss three supervised learning approaches; many others are outlined in Table 1.

A support vector machine (SVM) is a machine learning algorithm that finds the optimal hyperplane that separates classes of data. SVMs have been used extensively in the pharmaceutical industry for classification of small molecule data [124]. Depending on the kernel, training a SVM generally takes from $\mathcal{O}(N^2)$ to $\mathcal{O}(N^3)$. Rebentrost et al. [75] presented a quantum algorithm that can train a SVM with a polynomial kernel in $\mathcal{O}(\log N)$, and this was later extended to the radial basis function (RBF) kernel [76]. Unfortunately, it is unclear how to implement nonlinear operations, which are widely used in SVMs, given that quantum operations are restricted to be linear. On the other hand, a quantum computer enables other kinds of kernels that cannot be implemented in a classical computer [77].

Gaussian process (GP) regression is a method commonly used to build surrogate models, for example in Bayesian optimization [125]. GPs are also widely used to create quantitative structure–activity relationship (QSAR) models of drug properties [126], and recently also for molecular dynamics simulations [127]. One of the shortcomings of GP regression is the high cost ($\mathcal{O}(N^3)$) of inverting the covariance matrix. Zhao et al. [69] proposed the use of the HHL algorithm for linear systems to invert this matrix, achieving an exponential speedup as long as the matrix is sparse and well-conditioned, properties that are often achieved by covariance matrices. More importantly, this algorithm was extended to compute the logarithm of the marginal likelihood [70], which is a crucial step for hyperparameter optimization.

One of the most common methods in computational biology is the HMM, used widely in computational gene annotation and sequence alignment [59]. This method contains a series of *hidden* states, each of which is associated with a Markov chain; transitions between hidden states lead to changes in the underlying distribution. In principle, a HMM cannot be directly implemented in a quantum computer: sampling requires some kind of measurement, which would disrupt the system. However, there is a formulation in terms of *open* quantum systems—i.e., a quantum system that is in contact with the environment—that allows a Markovian system [78]. Although no improvements have been suggested to the classical

Baum–Welch algorithm to train a HMM, it has been found that quantum HMMs are more expressive: they can reproduce a distribution with fewer hidden states [79]. This may lead to richer applications across computational biology.

### 3.2.3 | Neural networks and deep learning

Recent developments in machine learning have been catalyzed by the discovery that multiple layers of artificial neural networks can discover intricate structures in raw data [108]. *Deep learning* has begun to permeate every scientific discipline, and in computational biology its successes include accurate contact prediction in proteins [61], improved diagnostics for several diseases [128], molecular design [129] and simulation [130,131]. Given the accomplishments of neural networks, there has been significant work to develop quantum analogues which can drive further improvements.

The name artificial neural network often refers to a multilayer perceptron, a neural network where every neuron takes a weighted linear combination of its inputs and returns the result through a nonlinear activation function. The major challenge in designing a quantum analogue is how to implement a nonlinear activation function employing linear quantum gates. There have been multiple recent proposals, and some ideas involve measurements [91–93], dissipative quantum gates [92], continuous-variable quantum computing [95], and the introduction of additional qubits to build linear gates that emulate nonlinearity [94]. These approaches aim to implement a *quantum* neural network, which is expected to be more expressive than a classical neural network due to the greater power of quantum information. The scaling advantage or disadvantage of training a multilayer perceptron in a quantum computer is unclear, although the focus has been in the possible increased expressivity of these models.

A vast amount of recent effort has focused on Boltzmann machines, recurrent neural networks able to act as generative models. Once trained, they can generate new samples that are similar to the training set. Generative models have important applications, for example in de novo molecular design [132,133]. Although Boltzmann machines are very powerful, computing the gradients to train them requires solving the NP-hard problem of sampling from a Boltzmann distribution, which has limited their practical application. Quantum algorithms have been proposed, using a D-Wave machine [101–103] or a circuit algorithm [104]; that sample from the Boltzmann distribution quadratically faster than it is possible classically [23]. Further work has presented a generalization, the quantum Boltzmann machine [134,135], which is expected to be more expressive. Recently, a heuristic to efficiently train quantum Boltzmann machines has been proposed that relies on system thermalization [105]. Moreover, some papers have proposed quantum implementations of generative adversarial networks, another very popular model [98–100]. These developments suggest an improvement in generative models as quantum computing hardware develops.

## 4 | EFFICIENT SIMULATION OF QUANTUM SYSTEMS

Chemistry is governed by the transfer of electrons. Chemical reactions, but also interactions between chemical entities, are also controlled by the distribution of electrons and the free energy landscape that these dictate. Problems like predicting whether a ligand binds to a protein, or understanding the catalytic pathway of an enzyme, reduce to understanding the electronic environment. Unfortunately, modeling these processes is extremely hard. The most efficient algorithm to compute the energy of a system of electrons, full configuration interaction (FCI), scales exponentially with the number of electrons [14], and even molecules with a handful of carbon atoms are barely accessible to computational investigation [136]. Although many approximate methods exist, including the deep and broad literature concerning density functional theory [137,138], they can be inaccurate and even fail dramatically in many situations of interest, like modeling the transition state of a reaction [139]. An accurate and efficient algorithm to study electronic structure would enable better understanding of biological processes, as well as open the door to next-generation engineering of biological interactions.

Quantum computers were initially suggested as a method to simulate quantum systems more efficiently [12,13]. In 1996, Seth Lloyd demonstrated that this is possible for some kinds of quantum systems [140], and a decade later Alán Aspuru-Guzik et al. showed that chemical systems are one of those cases [141]. For the past two decades, there has been significant research to fine-tune simulation methods for chemical systems that can calculate properties of interest. In the following section we provide a short overview—comprehensive reviews of quantum chemistry in a quantum computer have been authored by Cao et al. [26] and McArdle et al. [27].

## 4.1 | Advantages and shortcomings of quantum simulation

In principle, a quantum computer is able to efficiently solve the *fully correlated* electronic structure problem (the FCI equations), which would be a first step to accurately estimate binding energies, but also to simulate the dynamics of chemical systems. Classical computational chemistry has focused almost exclusively on approximate methods, which have been useful for example to estimate thermochemical magnitudes of small molecules [142], but that may not be adequate for processes involving bond cleavage or formation. By contrast, quantum processors can potentially solve the electronic problem by direct diagonalization of the FCI matrix, yielding the exact result within a particular basis set, and thus solving a myriad of problems arising from an incorrect description of the physics of molecular processes (e.g. ligand polarization). Moreover, unlike classical approaches, they do not necessarily require an iterative process, although as we will discuss the initial guess still plays an important role.

Although not as deeply explored, quantum computers also overcome limiting approximations that have been necessary in classical computing. For example, the real space formulation of quantum simulation automatically accounts for the nuclear wavefunction in the absence of the Born–Oppenheimer approximation [143]. This enables studying the nonadiabatic effects of some systems, which are known to be important in DNA mutation [144] and the mechanism of many enzymes [145]. Applications of quantum computing to relativistic simulations of systems have also been proposed [146], which are useful for the study of transition metals that appear in the active sites of many enzymes [147].

The work by Reiher et al. [148] provided some insight into the timescale of electronic structure calculations in a quantum computer. The authors considered the FeMo cofactor of the nitrogenase enzyme [148], whose nitrogen fixation mechanism is still not understood and is too complex to be studied with current computational approaches. A minimal basis FCI calculation on FeMoCo would require several months and circa 200 million of today's highest quality qubits. However, these estimations are bound to change with the rapid development of the field. In the 3 years since that publication, algorithmic advances have already lowered the time requirements by several orders of magnitude [149]. In addition to more powerful electronic structure methods, faster versions of current approximate methods that have been explored recently [150,151] may well accelerate prototyping, which would be of use for example in exploring reaction coordinates of enzymatic reactions, a problem that predates computational enzymology [152]. Moreover, through better understanding of intermolecular interactions, catalyzed by access to fully correlated calculations, or by access to faster throughput that improves parametrization, quantum simulations may well indirectly improve non-quantum simulation methods like force fields.

A final point to stress is that unlike other areas of algorithm research, like quantum machine learning, there are a number of near-term algorithms for quantum simulation that can be run in noisy, near-term devices. Multiple experimental groups around the world have reported successful demonstrations of these algorithms [66,153–156].

Unfortunately, there are also some disadvantages in quantum simulation of quantum systems. As discussed above, it is very difficult to extract information from a quantum computer. Although obtaining the energy is in general simple, retrieving the entire wavefunction is harder than computing the solution classically in the first place. This is an important drawback for chemical applications, where arguments based on the electronic structure are a powerful source of insight. In comparison with machine learning, however, the advantages vastly outweigh the shortcomings, and it is expected that quantum simulation will be one of the first useful applications of practical quantum computing [22].

## 4.2 | Algorithms for quantum simulation

### 4.2.1 | Fault-tolerant quantum computation

The quantum computers that can simulate large chemical systems will likely be fault-tolerant and therefore able to run arbitrarily deep algorithms without error. Such a quantum computer is able to simulate a chemical system by mapping the behavior of its electrons to the behavior of its qubits and quantum gates. The process for quantum simulation is conceptually very simple, and depicted in Figure 2a. We prepare a register of qubits that can store the wavefunction, and implement the unitary evolution of the Hamiltonian $e^{-iHt}$ with the Hamiltonian simulation techniques discussed below. With these elements, a quantum subroutine known as *quantum phase estimation* is able to find the eigenvectors and eigenvalues of the system. Namely, if the qubit register is initially in the $|0\rangle$ state, the final state will be:

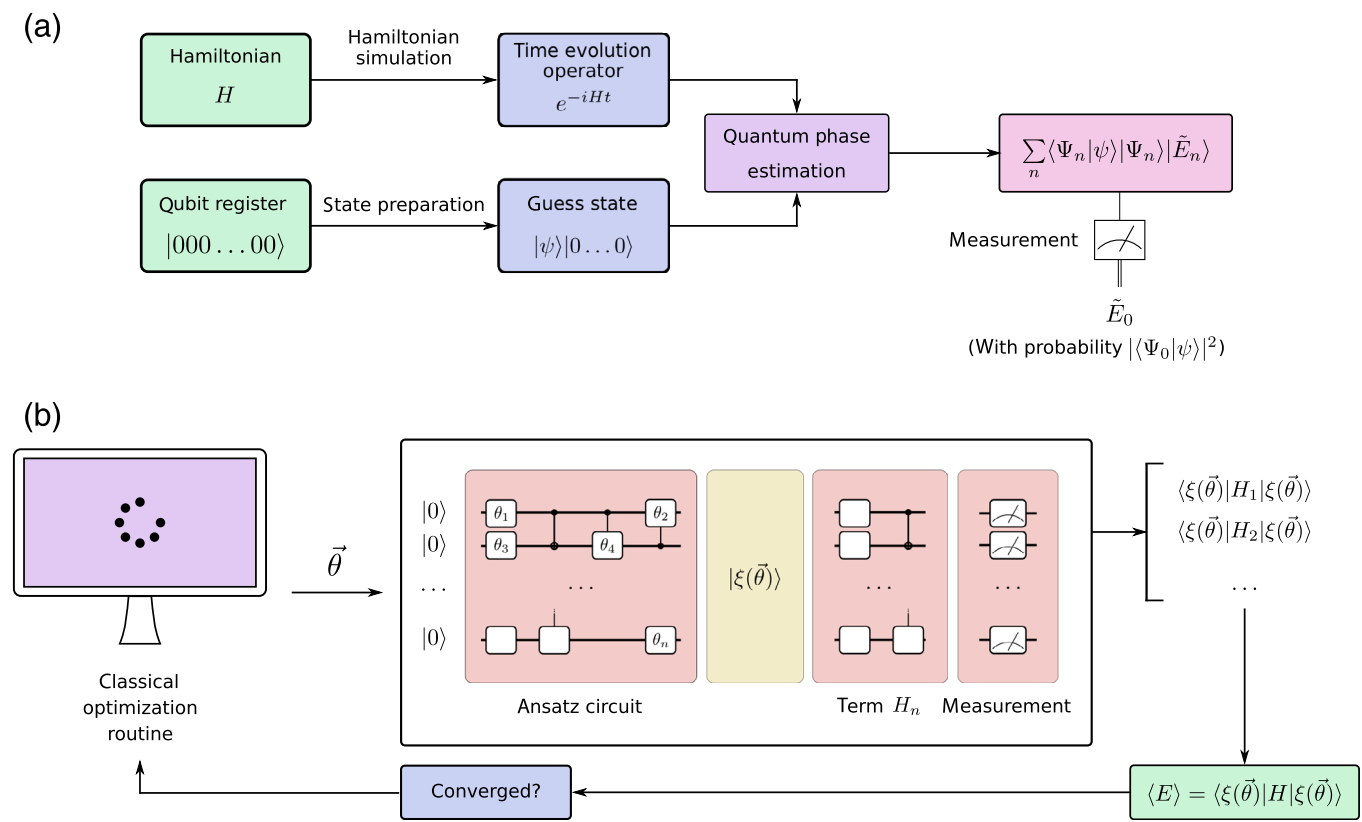

**FIGURE 2** (a) Algorithm for quantum simulation in a fault-tolerant quantum computer. The qubits are divided in two registers: one is prepared in a state $|\psi\rangle$ that resembles the objective wavefunction, while the other is left in the |0...0⟩ state. The quantum phase estimation (QPE) algorithm is used to find the eigenvalues of the time evolution operator $e^{-iHt}$, which is prepared using Hamiltonian simulation techniques. After QPE, a measurement of the quantum computer yields the energy of the ground state with probability $|\langle\Psi_0|\psi\rangle|^2$, hence the importance of preparing a guess state $|\psi\rangle$ with nonzero overlap with the true wavefunction. (b) Variational algorithm for quantum simulation in a near-term quantum computer. This algorithm combines the quantum processor with a classical optimization routine to perform multiple short runs that are quick enough to avoid errors. The quantum computer prepares a guess state $\left|\xi(\vec{\theta})\right\rangle$ with an *ansatz* quantum circuit dependent on several parameters $\{\theta_k\}$. The individual terms of the Hamiltonian are measured one by one (or in commuting groups, employing more advanced strategies), obtaining an estimation of the expected energy for a particular vector of parameters. The parameters are then optimized by the classical optimization routine until convergence. The variational approach has been extended to many algorithmic tasks beyond quantum simulation

$$\sum_{n=0}^{N}\langle\Psi_n|0\rangle\left|\tilde{E}_i\right\rangle|\Psi_n\rangle. \tag{7}$$

In other words, the final state is a superposition of the eigenvalues $\left|\tilde{E}_i\right\rangle$ and eigenvectors $|\eta_i\rangle$ of the system. The ground state is then measured with probability $|\langle\Psi_n|0\rangle|^2$. To maximize this probability, the initial state is set to be an easy to prepare, but also physically motivated state that is expected to be similar to the exact ground state. A common example is the Hartree–Fock state, although other ideas have been explored for highly correlated systems [157].

There are two common ways of representing the electrons in a molecule: grid-based and orbital or basis set methods (see McArdle et al. [27] for a full discussion). In basis set methods, the electronic wavefunction is represented as a sum of Slater determinants of electronic orbitals [158], which can be directly mapped to a qubit register [27,159,160]. This requires choosing a basis set and precomputing electronic integrals. On the other hand, in grid-based methods the problem is formulated as solving an ordinary differential equation in a grid. Grid-based simulations have the advantage of not requiring the Born–Oppenheimer approximation or a basis set. However, they are not naturally antisymmetric, which is required by the Pauli principle of quantum mechanics, so it is necessary to ensure antisymmetry using a sorting procedure [143,161]. Grid-based methods have been discussed in the context of simulating chemical dynamics [162] and computing the thermal rate constant [163]. Despite the differences, the workflow of quantum simulation is identical, as represented in Figure 2.

There are also several methods to build the operator $e^{-iHt}$. We will present the simplest technique, Trotterization, also known as the product formula approach [140]; for a full review see [26,27]. Trotterization is based on the premise that a molecular Hamiltonian can be split as a sum of terms describing one- and two-electron interactions. If this is true, then the operator $e^{-iHt}$ can be implemented in terms of the corresponding operators for every term in the Hamiltonian, using the Trotter–Suzuki formula [164]:

$$e^{-iHt} = \lim_{n \to \infty} \left( e^{-iH_1 t/n} \ldots e^{-iH_l t/n} \right)^n \approx \left( e^{-iH_1 t/N} \ldots e^{-iH_l t/N} \right)^N. \tag{8}$$

In second quantization, for example, every term in this expression will have the form $e^{-ih_{ij} a_i^\dagger a_j}$ or $e^{-ig_{ijkl} a_i^\dagger a_j^\dagger a_l a_k}$, where $a_i$ and $a_j^\dagger$ are respectively the annihilation and creation operators. Explicit, detailed constructions of the circuit representing these terms have been given by Whitfield et al. [165]. After computing the terms $h_{ij}$ and $g_{ijkl}$, known as *electronic integrals*, the $e^{-iHt}$ term is completely determined. It is also possible to use higher-order Trotter–Suzuki formulas to reduce the error. There are many other Hamiltonian simulation methods, an example of a powerful and sophisticated method being *qubitization* [166] and *quantum signal processing* [167], which have the provably optimal asymptotic scaling, although it is unclear whether this will translate to better practical applications.

### 4.2.2 | Near-term devices

Although fault-tolerance will be required for many applications, near-term quantum computers may be able to simulate small molecules beyond the reach of the best supercomputers [22]. Hybrid quantum-classical computation has been proved to be universal for quantum computing [168], which means that it can carry out any task a full quantum computer might do, even if with a significant overhead. An early hybrid quantum-classical algorithm was the *variational quantum eigensolver* (VQE), proposed and demonstrated by Peruzzo et al. [169] and later formalized by McClean et al. [170]. This algorithm, which underpins a great many other hybrid algorithms, is based on the Ritz-Rayleigh variational principle of quantum mechanics [14], which states that the expected energy of an approximate wavefunction will always be higher than the true ground state energy.

The algorithm employs an educated guess—an *ansatz*—dependent on several parameters $\vec{\theta}$. In a quantum processor, the ansatz is a circuit with parametrized quantum gates that acts on the input state for example, $|0\ldots0\rangle$ state. An example of a parametrized quantum gate is the $R_x(\phi) = \cos\left(\frac{\phi}{2}\right) \mathbb{I} - i\sin\left(\frac{\phi}{2}\right) X$ gate, which acts as a NOT gate for $\phi = \pi$ (up to a meaningless global phase that is not observable), as identity for $\phi = 0$, and adopts an intermediate behavior for different values of $\phi$. In effect the classical parameters $\vec{\theta}$ define a quantum state—namely, the state that the ansatz circuit will produce when configured to $\vec{\theta}$. The state may be impossible for a classical computer to describe or store explicitly. Using a technique known as *Hamiltonian averaging* [171], the energy of this wavefunction can be computed by a series of shallow circuits that can be run even in presence of errors. Then, a classical optimization subroutine is used to optimize $\vec{\theta}$. The variational principle ensures that the energy minimum is the best approximation to the wavefunction.

We recall that the expected value of an operator on a given state can be calculated as:

$$\begin{aligned}
\langle \psi | \hat{A} | \psi \rangle &= \langle \psi | \hat{A} \left| \sum_k c_k | \alpha_k \rangle \right\rangle = \sum_k c_k \langle \psi | \hat{A} | \alpha_k \rangle \\
&= \sum_{kk'} c_{k'}^* c_k \langle \alpha_{k'} | \hat{A} | \alpha_k \rangle = \sum_k |c_k|^2 a_k,
\end{aligned} \tag{9}$$

In this manipulation we have used the fact that the state $|\psi\rangle$ can be projected to the basis of eigenstates of $\hat{A}$, $\{\alpha_k\}$ [172]. We note that the coefficients $|c_k|^2$ represent the probability of yielding a given eigenstate upon measurement of the observable $\hat{A}$, hence Equation (9) is just the expected value of a probabilistic observation as understood in classical probability.

Hamiltonian averaging relies on the fact that a molecular Hamiltonian can be decomposed as the sum of angular momentum operators (i.e., tensor products of Pauli matrices) whose expectation values can be computed easily. After some manipulations, the expected value for the energy can be computed as:

$$\langle\psi|H|\psi\rangle = \langle E\rangle = \sum_{i_1\alpha_1} h^{i_1}_{\alpha_1}\left\langle\sigma^{i_1}_{\alpha_1}\right\rangle + \sum_{i_1\alpha_1} h^{i_1 i_2}_{\alpha_1\alpha_2}\left\langle\sigma^{i_1}_{\alpha_1}\sigma^{i_2}_{\alpha_2}\right\rangle + \dots \quad (10)$$

Every term in this expression is either known from the Hamiltonian expression, like the electronic integrals $h$, or can be obtained by sampling, adding a handful of quantum gates to the end of the ansatz and measuring the computational register, like the expected values of angular momentum operators. Each of these expected values requires a considerable number of repetitions to obtain accurate estimations, and this procedure must in turn be repeated for every term in the Hamiltonian. Given that this is a major cost in execution time, there has been significant research to make the procedure more efficient. For example, given that the expected value of a string of Pauli operators is contained in $[-1, +1]$, terms with small coefficients may be neglected [171], and more advanced truncation methods from classical quantum chemistry may also be employed [158]. Furthermore, it is possible to group commuting operators (e.g., single Pauli operators corresponding to different qubits) in the same variational burst [170].

A key point in algorithmic execution is the choice of a good ansatz. Although there is not, to date, a general theory of what constitutes a good ansatz, there have been several proposals that can be classified in two groups: physically motivated and hardware-efficient. Physically motivated ansätze are based on physically reasonable expressions of the electronic wavefunction. The archetypal example is the unitary coupled cluster ansatz, that is based on a unitary formulation of the popular coupled cluster method [14,173]. Hardware-efficient ansätze [153] are based on operations that are simple to implement on a particular device. While hardware-efficient ansätze are attractive due to their ease of implementation in current devices, recent work by McClean et al. has provided analytical and numerical evidence that random ansätze tend to lead to vanishing gradients when the number of qubits increases [174], meaning that it will be impossible to find a physically meaningful energy minimum. It is expected that physically motivated ansätze will not be affected by this inconvenience.

Although the VQE has dominated both the literature and the experimental explorations, there have also been proposals of different algorithms. Li and Benjamin proposed a variational simulator that considers the trajectory corresponding to the system's dynamics [57]. McArdle et al. extended this approach to imaginary time simulation, showing competitive scaling in comparison with gradient-based optimization in the VQE [175]. Other variational algorithms have been proposed for related simulation tasks, like preparing the Gibbs state, which allows computation of thermodynamic properties [176]. These and other methods try to achieve the best efficiency with minimum resources, in order to leverage the quantum computers available during the next decade.

## 5 | OPTIMIZATION PROBLEMS

Many problems in computational biology and elsewhere can be formulated as finding the global minimum or maximum of a complicated, high-dimensional function. For example, it is believed that the native structure of a protein is the global minimum of its free energy hypersurface [177]. In a different area, determining a community in a network of interacting proteins or biological entities is equivalent to finding an optimal subset of the nodes [178]. Unfortunately, with the exception of a few simple systems, optimization problems are often very difficult, even NP-complete or NP-hard. Although there exist heuristics to find approximate solutions, these tend to provide only local minima, and in many cases, even the heuristics are intractable. The ability of quantum computers to accelerate such optimization problems or find better solutions has been explored in depth.

The subject of optimization in a quantum computer is complicated, since it is often not obvious whether a quantum computer can provide any kind of speedup. In this section, we will provide an overview of some ideas in quantum optimization. Although the guarantees for improvement are not as clear as in, for example, quantum simulation, it is expected that in the long-term these approaches will be useful.

### 5.1 | Optimization in a quantum processor

Adiabatic quantum optimization is one of the most popular approaches in optimization, spearheaded by the availability of D-Wave machines [179,180] that implement this approach natively. Adiabatic quantum computing [181] is based on the adiabatic theorem of quantum mechanics [182]. According to this theorem, if a system is prepared in the ground state of a Hamiltonian, and this Hamiltonian is varied slowly enough, the system will always remain in its instantaneous ground state. This may be exploited to perform computation, by encoding a problem (e.g., a score function that we hope to minimize) as a Hamiltonian, and slowly evolving towards this Hamiltonian from some starting system that can be trivially prepared in its ground state. In general, adiabatic evolution is expressed as:

$$H(t) = \mathcal{A}(t) H_{\text{initial}} + \mathcal{B}(t) H_{\text{final}}. \quad (11)$$

Here $\mathcal{A}(t)$ and $\mathcal{B}(t)$ are functions such that $\mathcal{A}(0) = \mathcal{B}(T) = 1$ and $\mathcal{A}(T) = \mathcal{B}(0) = 0$ for certain runtime $T$. For example, we could consider a linear annealing program with $\mathcal{A}(t) = (1 - t/T)$ and $\mathcal{B}(t) = t/T$. There has been significant discussion about the runtime of the adiabatic algorithm, but a common heuristic is that the runtime is at most proportional to the inverse square of the minimum spectral gap (the smallest energy difference between the ground and first excited states) during adiabatic evolution $\mathcal{O}(1/\Delta^2)$. Unfortunately, many interesting problems have been found with an exponentially vanishing gap with increasing problem sizes [183–185]. It is believed that adiabatic quantum computing (and quantum computing, in general) is unable to efficiently solve the class of NP-complete problems, or at least no such feat has withstood scrutiny [17,186].

In principle, adiabatic quantum computing is equivalent to universal quantum computing [187]. This universality holds only if the evolution is allowed to be nonstoquastic, meaning that the Hamiltonian has nonnegative off-diagonal elements at some point of the evolution. The most popular experimental implementation of adiabatic quantum computing, commercialized by the company D-Wave Systems Inc., employs stoquastic Hamiltonians [180], and it is therefore nonuniversal. There are some concerns in the literature that this variety of quantum computing may be classically simulatable [17], meaning that an exponential speedup may be impossible. Despite these concerns, it has been used widely as an optimization metaheuristic and has recently been shown to overperform simulated annealing [188].

Quantum optimization has been studied beyond the adiabatic model. The quantum approximate optimization algorithm (QAOA) [189–191] is a variational algorithm for optimization in a near-term quantum computer that has gathered considerable interest in the literature. There have been several experimental implementations of QAOA in quantum processors, for example, [192] Figure 3.

## 5.2 | Protein structure prediction

Template-free protein structure prediction is still a major open problem in computational biology. The solution to this problem has extensive applications in molecular engineering and drug discovery. According to the funnel hypothesis of protein folding, the native structure of a protein is believed to the global minimum of its free energy [177,193], although many counterexamples exist. Given the vast conformational space available to even small peptides, exhaustive classical simulations are intractable. However, many have wondered if quantum computing may be able to assist this problem.

The quantum computing literature has focused on the protein lattice model, where the peptide is modeled as a self-avoiding walk on a lattice, although other models have started to appear [194]. Every node of the lattice corresponds to a residue, and the interactions between spatial neighbors contribute to an energy function. There are several energy contact schemes, but only two have been used in quantum applications: the hydrophobic-polar model [195] which considers only two classes of amino acids, and the Miyazawa–Jernigan [196] containing interactions for every pair of residues. Although these models are notable simplifications, they have provided significant insight into the principles of protein folding [197], and have been proposed as a coarse-grained proxy to explore conformational space before further refinement [198,199].

Nearly all work has centered on adiabatic quantum computing, since even toy peptides require a large number of qubits, and D-Wave quantum annealers are the largest quantum devices available so far. However, a recent article by Fingerhuth et al. [200] has also attempted to use the QAOA algorithm. Both methods share similar characteristics once the protein lattice problem is encoded as a Hamiltonian operator. This method was first addressed by Perdomo et al. [201], who proposed the use of a register of $DN\log_2 N$ qubits to encode the Cartesian coordinates of $N$ amino acids on a $D$-dimensional cubic lattice of side $N$. The energy function is expressed in the Hamiltonian, containing terms to reward protein contacts, preserve primary structure and avoid amino acid overlaps. Soon after this landmark article, another paper by Babbush et al. discussed the construction of more bit-efficient models for a 2D lattice [202].

These encodings were tested in a real device in 2012 when Perdomo et al. [203] computed the lowest-energy conformation of the PSVKMA peptide on a D-Wave quantum annealer. Recently, Babej et al. extended the "turn" and

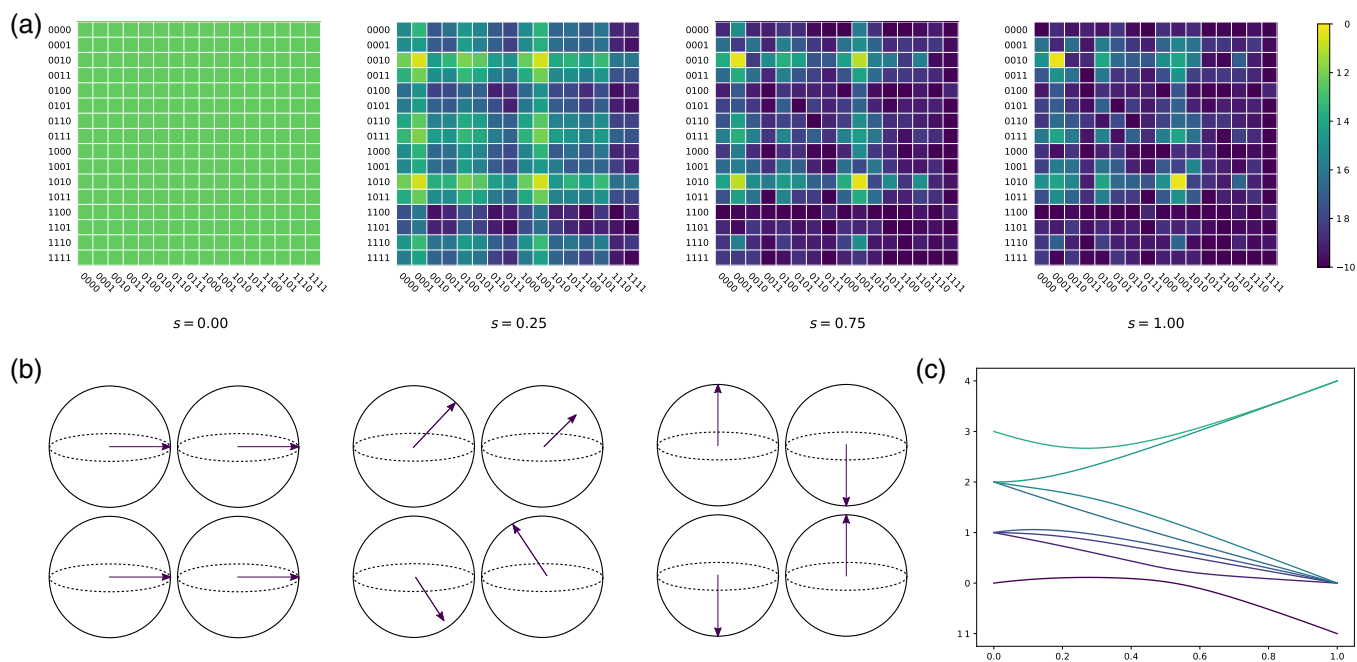

**FIGURE 3** (a) Simulation of an adiabatic quantum computer implementing a simplified protein folding problem, described in [201]. Color encodes the decimal log probability of a particular binary string. At the end of the computation, the two lowest-energy solutions have a probability of measurement close to 0.5. In finite time, the evolution is never entirely adiabatic, and other binary strings have residual probabilities of measurement. (b) Depiction of the adiabatic quantum computing process. The potential governing the qubits is slowly modified, causing them to rotate. Note that the Bloch sphere representation is incomplete, as it does not depict the correlations between different qubits, which are necessary for quantum advantage. At the end of the evolution, the system of qubits is in a classical state (or a superposition of classical states) representing the lowest-energy solution. (c) Energy levels during the adiabatic quantum evolution. When other levels are close to the ground state, the population at the ground state can leak towards excited states. The amount of time required to ensure quasi-adiabatic evolution is governed by the minimum energy difference between the levels $\Delta$, which is indicated by a dotted line

"diamond" encodings to 3D models, and introduced algorithmic improvements that reduce the complexity and $k$-locality of the "turn" encoding Hamiltonians [204]. Their work employed the D-Wave 2000Q processor to determine the ground state of chignolin (10 residues) on a square lattice and Trp-Cage (8 residues) on a cubic lattice, which are the largest peptides explored to date. These experimental implementations use the *divide-and-conquer* technique, in which a portion of the peptide is fixed. This allows the introduction of larger problems into the quantum computer, at the cost of examining $2^N$ more problems.

Finding the lowest-energy conformation of a lattice model is an NP-hard problem [205,206], meaning that under standard hypotheses, no polynomial-time classical algorithm for this problem exists. Furthermore, it is currently believed that quantum computers cannot offer an exponential speedup to NP-complete and harder problems [207], although they can offer scaling advantages that have been known in the literature as "limited quantum speedup" [208]. A recent study by Outeiral et al. has employed numerical simulations to investigate this fact, concluding that there is evidence for a limited quantum speedup, although this result may require adiabatic machines using error correction or quantum simulation in fault-tolerant universal machines [209].

Although most of the literature has focused on the protein lattice model, a recent article [210] has attempted to use quantum annealing to perform rotamer sampling in the Rosetta energy function [211]. The authors used the D-Wave 2000Q processor finding a scaling that seemed almost constant in comparison with classical simulated annealing. A very similar approach was presented by Marchand et al. [212] for conformer sampling.

## 6 | CONCLUSIONS

A quantum computer can store and manipulate states which, if represented classically, would correspond to vast amounts of information, and execute some algorithms that are exponentially faster than any classical alternative. The potential of even small quantum computers to outperform the best supercomputers on certain tasks may prove transformative to computational biology, promising to make impossible problems difficult, and difficult problems routine. Early quantum processors that can solve useful problems are expected to arise within the next decade. Understanding what quantum computers can and cannot do is therefore a priority for every computational scientist. In this review, we have attempted to give an overview of the technology that may soon revolutionize computational biology. Although we have been unable to cover every topic, we have introduced the principles of quantum information processing and discussed three areas of importance.

Although we are just entering the era of practical quantum computation, it is already possible to glimpse an emerging quantum computational biology within the next decades [29]. We hope that this review will arouse the interest of the computational biologist in the technologies that may soon disrupt their field, and attract the quantum information practitioner to the fertile field of computational biology, from where many meaningful applications are expected to arise.

### ACKNOWLEDGMENTS
The authors would like to thank Anne Nierobisch and Susan Leung for useful discussions, and Daniel Nissley, Niel de Beaudrap and Sam McArdle for enlightening comments on an early version of this manuscript. C. O. would like to thank F. Hoffmann-La Roche, UCB and the UK National Quantum Technologies Programme for financial support through an EPSRC studentship (EP/M013243/1). G. M. M. thanks the EPSRC and MRC for support via EP/L016044/1 and EP/S024093/1. S. C. B. acknowledges support from the EU Flagship project AQTION, the NQIT Hub (EP/M013243/1) and the QCS Hub (EP/T001062/1).

### CONFLICT OF INTEREST
The authors have declared no conflicts of interest for this article.

### AUTHOR CONTRIBUTIONS
**Carlos Outeiral:** Conceptualization; writing-original draft; writing-review and editing. **Martin Strahm:** Writing-review and editing. **Jiye Shi:** Writing-review and editing. **Garrett Morris:** Writing-review and editing. **Simon Benjamin:** Writing-review and editing. **Charlotte Deane:** Conceptualization; writing-original draft; writing-review and editing.

### ORCID
*Carlos Outeiral* https://orcid.org/0000-0003-1408-5554
*Jiye Shi* https://orcid.org/0000-0002-9628-8680
*Garrett M. Morris* https://orcid.org/0000-0003-1731-8405
*Simon C. Benjamin* https://orcid.org/0000-0002-7766-5348
*Charlotte M. Deane* https://orcid.org/0000-0003-1388-2252

### RELATED WIREs ARTICLES

Coarse-grained models of protein folding as detailed tools to connect with experiments

Biomolecular simulations: From dynamics and mechanisms to computational assays of biological activity

Formatting biological big data for modern machine learning in drug discovery

Global optimization